\documentclass[a4paper,fleqn,usenatbib,usedcolumn]{mnras}

\usepackage{newtxtext,newtxmath}

\usepackage[T1]{fontenc}
\usepackage{ae,aecompl}

\usepackage{graphicx}	
\usepackage{amsmath}	
\usepackage{amssymb}	
\usepackage{multirow}
\usepackage{cases}
\usepackage{float}

\hypersetup{pdfauthor={W. N. Alston},
               pdftitle={Non-stationary variability in accreting compact objects},
               pdfkeywords={accretion, accretion discs -- X-rays: binaries -- X-rays: galaxies -- galaxies: individual: IRAS 13224--3809},
               bookmarksnumbered=true}

   \volume{{\rm in press}}
   
\def\apj{ApJ}
\def\mnras{MNRAS}
\def\apjl{ApJ}

\def\aap{A\&A}

\def\apjs{ApJS}

\def\Msun{\hbox{$\thinspace M_{\odot}$}}

\def\xmm{{\it XMM-Newton~\/}}

\newcommand{\xmmn}{{\it XMM-Newton~\/}}

\def\Mbh{\hbox{$M_{\rmn{BH}}$}}

\newcommand{\iras}{IRAS 13224--3809~\/}
\newcommand{\irasno}{IRAS 13224--3809}

\defcitealias{alston19a}{A19}
\defcitealias{uttleymchardyvaughan05}{UMV05}
\defcitealias{uttleymchardyvaughan17}{UMV17}

\def\gsim{\mathrel{\hbox{\rlap{\hbox{\lower4pt\hbox{$\sim$}}}\hbox{$>$}}}}
\def\lsim{\mathrel{\hbox{\rlap{\hbox{\lower4pt\hbox{$\sim$}}}\hbox{$<$}}}}

\title[Non-stationary variability]{Non-stationary variability in accreting compact objects}

\author[W. N. Alston]{
W. N. Alston,$^{1}$\thanks{E-mail: wna@ast.cam.ac.uk}
\\
$^{1}$Institute of Astronomy, Madingley Rd, Cambridge, CB3 0HA.
}

\date{Accepted 2019 February 07. Received 2019 February 05; in original form 2019 January 31}

\pubyear{2019}

\begin{document}
\label{firstpage}
\pagerange{\pageref{firstpage}--\pageref{lastpage}}
\maketitle

\begin{abstract}
Accreting compact objects show variations in source flux over a broad range of timescales and in all wavebands.  The light curves typically show a lognormal distribution of flux and a linear relation between flux and rms.  It has been demonstrated that an exponential transform of an underlying (and unobserved) Gaussian stochastic process provides a very good description of the light curves with these observed properties.  Recently, a non-stationary power spectrum was observed on fast timescales ($\sim$\,days) in the active galaxy, IRAS 13224--3809, as well as a non-lognormal flux distribution and non-linear rms-flux relation.  Here, we investigate the affects of \emph{piecewise} non-stationary power spectra on the resultant flux distribution and rms-flux relation.  We demonstrate that the simple ``exponentiation'' model successfully reproduces the observed quantities, even when the light curves are non-stationary.  We also demonstrate how non-lognormal flux distributions and rms-flux relations inconsistent with a linear model can be erroneously produced from poorly sampled PSDs.  This is of particular importance for AGN surveys where very long baselines are required to sample the PSD down to low enough frequencies.

\end{abstract}




\section{Introduction}


One of the defining characteristics of accreting compact objects is their variation in source flux, observed in many wavebands and over a broad range of timescales.  The variability is a stochastic process which can be described using a power spectral density (PSD), typically showing broadband noise power over several decades in frequency.  The origin of the variability and origin of emission mechanisms are not necessarily the same thing.  It is therefore crucial to understand the properties of the variability process if we want to i) understand the underlying accretion physics, and ii) use the variability to perform spectral timing studies to infer important properties of the accretion flow.

Another apparent universal feature of accreting objects is the linear relationship between the short-term rms amplitude and mean flux variations on longer time-scales.  Known as the linear rms-flux relation (\citealt{UttleyMchardy01}), this has been observed in active galactic nuclei (AGN, e.g. \citealt{vaughan11a}), neutron star (NS) and black hole (BH) X-ray binaries (XRBs, e.g.  \citealt{gleissner04}; \citealt{HeilVaughanUttley12}), ultraluminous X-ray sources (ULXs, \citealt{HeilVaughan10}; \citealt{hernandezgarcia15}), as well as the fast optical variability from XRBs (\citealt{Gandhi09}), blazars (\citealt{edelson13}), accreting white dwarfs (\citealt{scaringi12}; \citealt{dobrotka15};\citealt{vandesande15}) and young stellar objects (\citealt{scaringi15}).  Crucially, a given source exhibits a linear relation over a broad range of timescales. 

Models for the variability process often only explain the shape of the PSD.  However, this is not the best tool to distinguish among models, as many types of models can produce identical PSD shapes.  The lognormal flux distribution, rms-flux relation and the resulting non-linear variability appear to be more fundamental features of the underlying process.  These characteristics rule out many types of models to describe the process, including additive shot-noise, self-organised criticality and any model which consist of multiple additive independent varying regions.  The process is required to be multiplicative in order to produce the observed variability characteristics.

\citet[][hereafter UMV05]{uttleymchardyvaughan05} proposed the simple ``exponentiation" model to explain the key observed light curve properties in a single energy band.  Here, an underlying Gaussian distributed light curve is exponentiated to produce a non-linear light curve with the same properties as the data: a lognormal distribution in fluxes as well as a linear rms-flux relation.  Importantly, the model produces the desired linear rms-flux relation over all timescales.

It is important to distinguish here models for the variability process with models for the underlying physics producing the observed variability.  The leading physical model is based on the inward propagation of random accretion rate fluctuations in the accretion flow (e.g. \citealt{Lyubarskii97}; \citealt{kotov01}; \citealt{king04}; \citealt{zdziarski05}; \citealt{arevalouttley06}; \citealt{ingramdone10}; \citealt{kelly11}, \citealt{IngramvanderKlis13}; \citealt{Scaringi14}.  In this model, slow fluctuations in the local mass accretion rate at large radii propagate through the accretion flow and modulate faster fluctuations at smaller radii.  This multiplicative model inherently produces variations coupled together over a broad range in timescales.  Numerical accretion disc simulations (e.g. \citealt{CowperthwaiteReynolds14}) including the affects of magnetohydrodyamics (MHD; \citealt{HoggReynolds16}) are starting to reproduce the observed flux distribution and rms-flux relation.


Another important property of a stochastic process is the concept of stationarity.  A process is said to be \emph{strongly} stationary when its mean and variance tend to some well defined value on long timescales.  In the case of accreting objects, this `long' timescale should at least correspond to the timescale required to measure the low-frequency rollover in the PSD. For AGN, observing constraints mean only the red noise part of the PSD at high frequencies is typically observed.  They are therefore considered only \emph{weakly} stationary, as the low frequency break in the PSD is typically not observed: this roll-over to zero amplitude provides a well defined mean at some finite duration.  The full range of light curve behaviour is therefore not observed in a typical observation.  A process is considered non-stationary when one or more of its moments do not satisfy these conditions.  The statistics literature has a strict definition of stationarity, whereas time series can be non-stationary in an infinite number of ways.

In \citet[][hereafter A19]{alston19a} we revealed non-stationary variability on very fast timescales ($\lsim$\,days) in the X-ray light curves of the AGN, \irasno.  The non-stationarity in \iras manifested itself as changes in the PSD shape and normalisation with time, in addition to the changes in mean flux over any given segment the PSD is measured.  With a central black hole mass of $\Mbh \sim 10^{6} \Msun$, this is equivalent to the broadband noise PSD changing after just $\sim 10$\,s in a $10 \Msun$ BH XRB.

In addition to the strong non-stationarity, \iras displays a non-lognormal distribution of source flux, as well as a non-linear rms-flux relation on all observed timescales.  The relation was well modelled as $\mathrm{rms} \propto \mathrm{flux}^{2/3}$, meaning the source is fractionally more variable at lower source flux, as was evident in the PSD.  Given the short duration the deviations from the expected observables occur on, an important question to ask is whether the simple exponentiation model of \citetalias{uttleymchardyvaughan05} still work in the case of non-stationary light curves, or is some other more complex process model required to produce the observed quantities?

In this paper we explore the effects of non-stationarity light curves with the exponentiation model and use this to describe the observed variability properties of \irasno.  We will also demonstrate the affects of observational constraints on the true PSD for the measured flux distributions and rms-flux relations.  This has big implications when using the variability properties for understanding the accretion process, in particular from AGN time domain surveys.




\section{Results}
\label{sec:results}

\begin{figure*}
	\includegraphics[width=0.48\textwidth,angle=90]{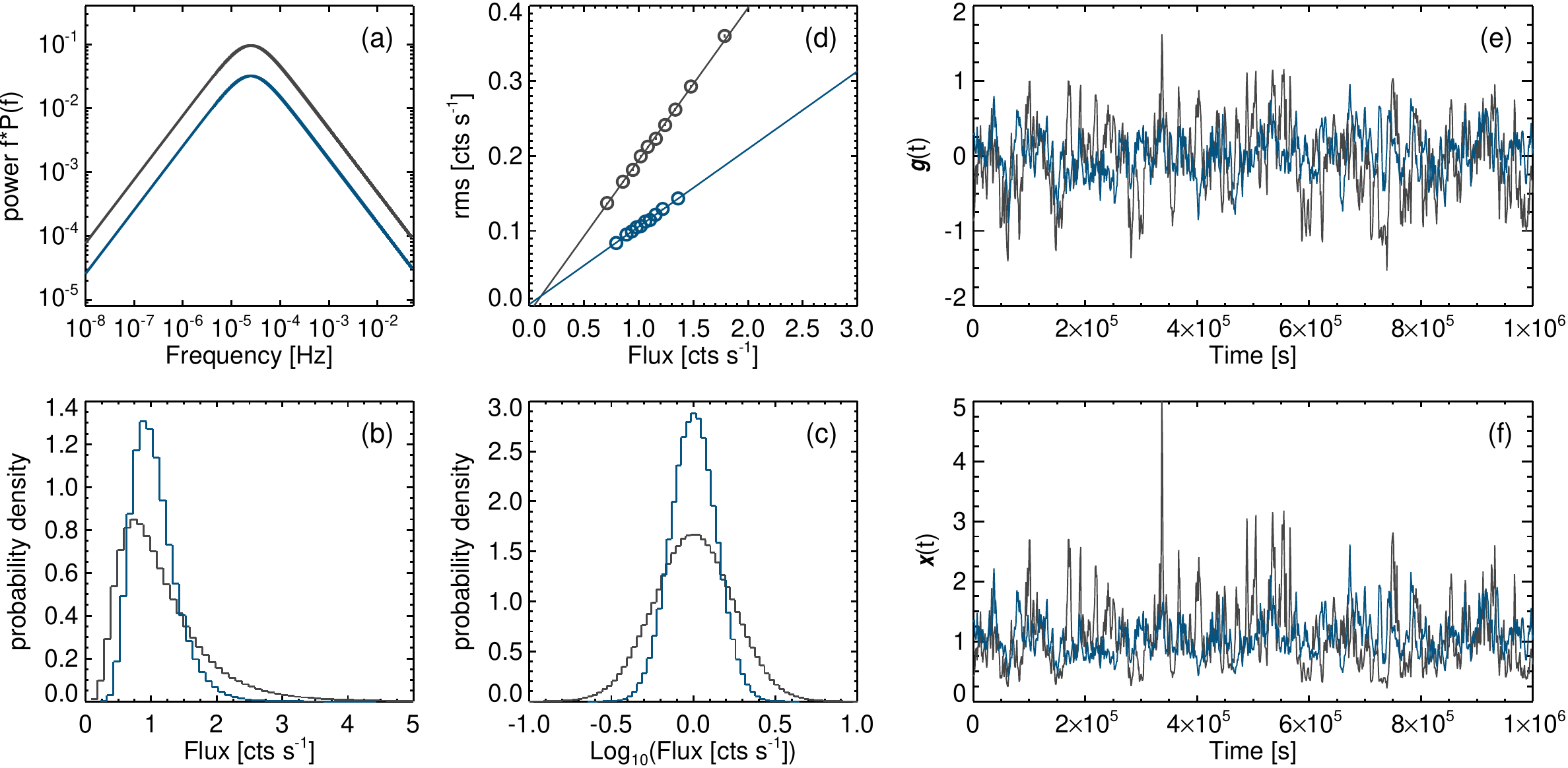}
    \caption{Demonstration of the exponentiation model for producing stationary light curves with the properties seen in accreting compact objects.  Panel (a) shows model PSDs formed of identical broad ($q = \nu_{\rm c} / \sigma_{\rm L} = 0.2$) Lorentzians with normalisations differing by a factor of 3.  The resulting exponentiated light curve flux distributions are shown in panel (b) and the log transformed fluxes in panel (c).  The $0.2-0.1$\,mHz rms-flux relations for each model PSD are shown in panel (d). Panel (e) shows the zero mean Gaussian distributed stationary light curve $g(t)$.  Panel (f) shows the light curve $x(t) = {\rm exp}[g(t)]$ after being scaled to a mean of unity and variance of the model PSD.}
    \label{fig:example}
\end{figure*}

\begin{figure*}
\includegraphics[width=0.54\textwidth,angle=90]{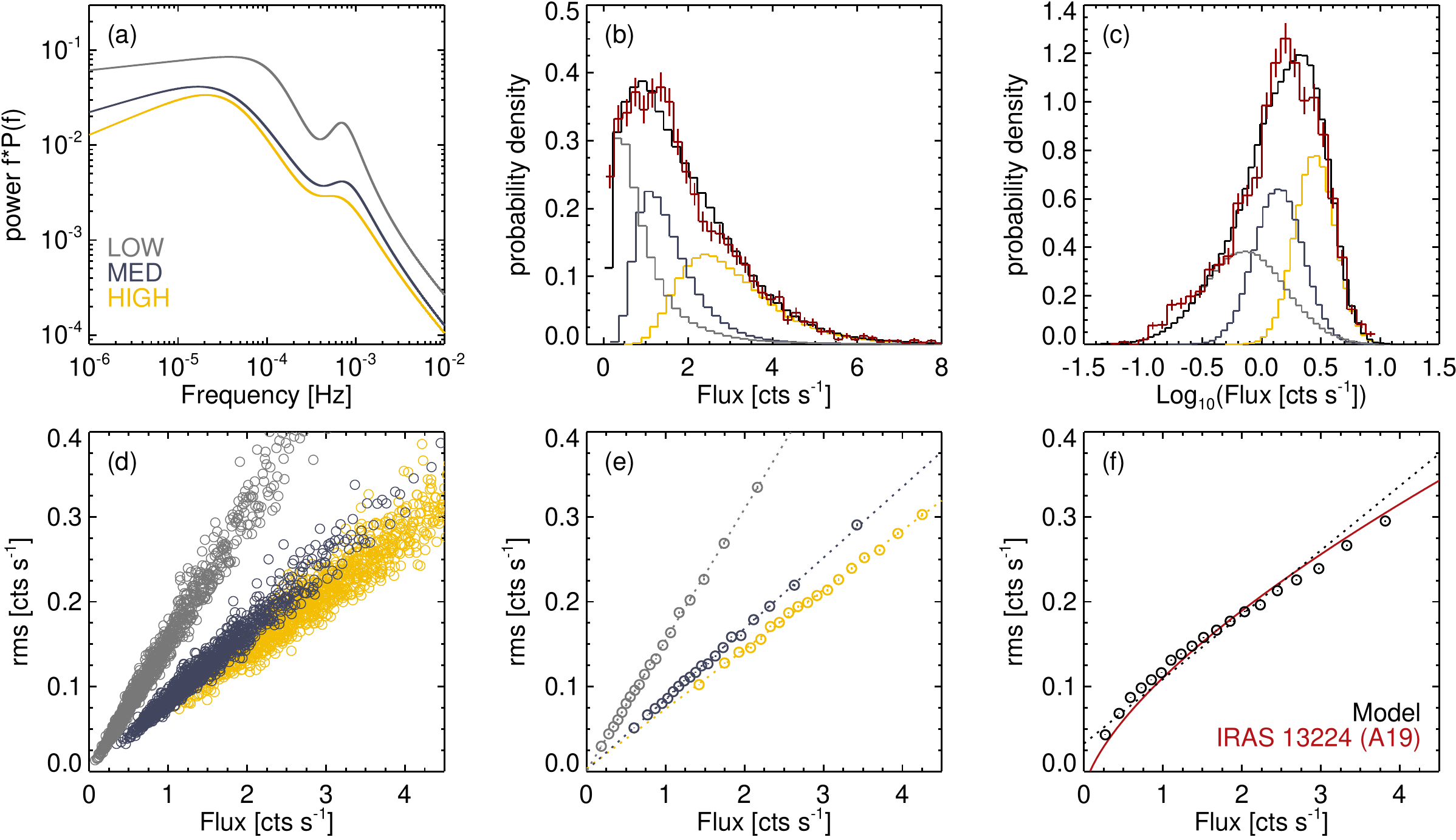}
    \caption{The exponentiation model applied to the piecewise non-stationary PSDs observed in \irasno.  Panel (a) shows the rms normalised model PSDs for the low (grey), medium (blue) and high (yellow) flux bins.  Panel (b) shows the flux distributions of the model light curves, shown as a probability density (normalised to unity).  The black curve is the total model light curve and the red curve is the \iras data from \citetalias{alston19a}.  Panel (c) shows the log transformed flux distributions.  Panel (d) is the unbinned $0.2-0.1$\,mHz rms-flux relation for each model light curve.  Panel (e) shows the same data after rebinning to $100$ estimates per flux bin.  The dotted lines are the best-fitting linear models.  Panel (f) shows the rms-flux relation for the total model light curve.  The red curve is the best fitting model from \citetalias{alston19a} and the black dotted line is the best fitting linear model.}
    \label{fig:iras}
\end{figure*}

\subsection{The exponentiation model}
\label{sec:expo}

We start by demonstrating the exponentiation model using some simple broadband PSDs and inspecting the resulting flux distributions and rms-flux relations.  Light curves are simulated following the approach described in detail in \citetalias{uttleymchardyvaughan05} and \citet[][hereafter UMV17]{uttleymchardyvaughan17}.  The model produces light curves $x(t)$ with a linear rms-flux relation on all timescales by taking the exponential of a linear and stationary aperiodic light curve.  We used the \citet{timmerkonig95} method to produce a zero-mean Gaussian distributed stationary light curve, $g(t)$.  The model light curve is then generated by $x(t) = {\rm exp}[g(t)]$, which is non-linear by design.  We simulate (Poisson noise free) light curves of length $T = 1 \times 10^{8}$\,s and time resolution ${\rm dt}=1$\,s.  The non-linear transformation increases the total variance of the light curves, so these are then scaled to the variance of the input PSDs (\citetalias{uttleymchardyvaughan05}).  The light curves were then rebinned to ${\rm dt}=50$\,s to alleviate aliasing effects (e.g. \citealt{Uttley02a}; \citealt{alston13a}).  


Fig.~\ref{fig:example} (a) shows two broad PSD models formed from Lorentzian functions:

\begin{equation}
\label{eqn:lorentz}
   P(\nu) = \frac{N_{\rm L} (\sigma_{\rm L} / 2 \pi) }{(\nu - {\nu}_{\rm c})^2 + (\sigma_{\rm L} / 2)^2}
\end{equation}

\noindent with coherence $q = 0.2$ (where $q = \nu_{\rm c} / \sigma_{\rm L}$, with centroid $\nu_{\rm c}$ and width $\sigma_{\rm L}$).  In this example the Lorentzian normalisation differs between the two models by a factor of 3.  The model light curves $g(t)$ are shown in Fig.~\ref{fig:example} panel (e).  Following the exponentiation, the model light curves are scaled to a mean of unity, with $x(t)$ shown in panel (f).  The simulated light curve flux distribution is shown in panel (b), which are well modelled by a three-parameter lognormal model.  The log-transformed fluxes are shown in panel (c). It is obvious from these plots that the skewness of the light curves increases with fractional rms (see also equation 15 in \citetalias{uttleymchardyvaughan05}). 

We determine the rms-flux relation of the simulated light curves following the procedure described in \citet{vaughan11a}, \citet{HeilVaughanUttley12} and \citetalias{alston19a}.  The simulated light curves are split into segments with length $T_{\rm seg} = 10^{5}$\,s, resulting in an rms estimate for each segment.  The data points are then sorted on mean flux and binned to have $> 50$ estimates per bin.  The $0.2-0.1$\,mHz rms-flux relations are shown in Fig.~\ref{fig:example} panel (d), which as expected are well described by a linear model.  As well as having a larger spread in flux the model light curve with higher variance also has a larger spread in rms values.  As the mean rate of both model light curves is identical, the model with larger variability power is fractionally more variable, resulting in an rms-flux relation with a steeper gradient.

\subsection{A non-stationary model for \iras}
\label{sec:iras}

We now extend this simple exponentiation model to describe the non-stationary light curves in \iras \citetalias{alston19a}.  This manifested itself as changes in the PSD shape and normalisation with time, in addition to the changes in mean flux over any given segment the PSD is measured.  For any given time series, one of the difficulties in measuring any changes in the PSD with time reliably, is the trade-off between variances of the estimates (i.e. error bars) and the frequency resolution of the data (i.e. the lowest frequency the PSD can be measured to).  In \irasno, several neighbouring \xmmn observations were required to provide enough signal-to-noise to show a statistically significant change in the PSD.  The light curves in \iras can be described as \textit{piecewise} non-stationary; within the epoch used to measure a PSD it can be treated as stationary as the changes in the PSD are too small to be significantly measured.  For simplicity we use only the $0.3 - 1.2$\,keV (soft) band due to the strong energy dependence of the variability.  The $1.2 - 10.0$\,keV (hard) band also shows the same non-stationary behaviour (\citetalias{alston19a}).

We use the PSD fits from \citetalias{alston19a} (Fig. 10) where the data are sorted into 3 flux bins of equal duration; the PSDs within a flux bin are statistically equivalent.  The PSDs consist of two broad Lorentzian components; one peaking at $\sim 10^{-5}$\,Hz and one at $\sim 10^{-3}$\,Hz.  One of the important findings of \citetalias{alston19a} was that the low-frequency component rolls over to a slope $\alpha < 1$ down to low frequencies; we are seeing the bulk of the variability power, which is all occurring on timescales measurable with current observations.  As the source flux decreases the low frequency component moves to higher frequency ($\sim 10^{-4}$\,Hz).  The centroid frequency, $\nu_{\rm c}$ of the high frequency component remains approximately constant.

The model PSDs with rms normalisation are shown in Fig.~\ref{fig:iras} (a), where it can be seen how the data are fractionally more variable at lower source flux (\citetalias{alston19a}).  In a similar manner to the previous section, we simulate (Poisson noise free) light curves of length $T = 1 \times 10^{7}$\,s and time resolution ${\rm dt}=1$\,s from each of the 3 PSDs.  Following the exponentiation the light curves are rebinned to a time resolution of ${\rm dt}=50$\,s.  The model light curves were then scaled to the observed count rate of $1.0, 1.6$ and $3.1$\,${\rm cts\,s^{-1}}$ for the low, medium and high flux bins respectively.  The light curves are then scaled to the variance of the respective input PSDs.  The total model light curve is then the concatenation of the piecewise stationary light curves, producing a piecewise non-stationary light curve.

Fig.~\ref{fig:iras} (b) shows the flux distributions for the 3 model light curves as well as their sum, representing what is captured in the `real' observations.  As expected from the results of the previous section, the individual flux distributions are lognormal (or Gaussian when log transformed).  When the sum of all three pieces is captured, this produces a clear deviation from a lognormal relation.  This is due to the mean values of each distribution differing as well as the variances.  Also shown in Fig.~\ref{fig:iras} (b) is the observed flux distribution from \citetalias{alston19a}, with error bars proportional to $\sqrt{N}$ data points per histogram bin.  The model light curves are a remarkably good fit to the data, with the skew to low fluxes being captured by the model.  There is slight disagreement with the data, but this is to be expected as the exponentiation is a very simplified model (\citetalias{uttleymchardyvaughan05}; \citetalias{uttleymchardyvaughan17}) and the simulation is one realisation of a stochastic process.

The rms-flux relation is determined in a similar manner to before.  In Fig.~\ref{fig:iras} panel (d) we show the unbinned rms-flux relation for each piecewise model simulation.  The spread in values for each piecewise model can be seen, as well as the overlap in estimates from the individual model; the flux binning to form the relation is performed on each piecewise model light curve only.  Panel (e) shows the binned ($> 50$ estimates) rms-flux estimates for each piecewise model light curve.  As individual models are stationary, they produce a linear rms-flux relation following the exponentiation, with the gradient of each relation related to the variance of the model light curve.   

If we treat the 3 model light curves as one large model, we can perform the flux sorting and binning on the model as a whole, representing what is typically done with real data.  Fig.~\ref{fig:iras} panel (f) shows the rms-flux relation for the total model light curves with $>100$ estimates per bin.  The resulting relation shows a clear deviation to a linear model, with a linear model ruled out with $p$-value $< 10^{-5}$.  Overplotted in panel (f) is the best fit model with $\mathrm{rms} \propto \mathrm{flux}^{2/3}$ to the $0.2-0.1$\,mHz data from \citetalias{alston19a}.  We note here that the observed non-linear rms-flux relations at other frequency bands are also reproduced by the model.  As before, the model is not a perfect fit to the data, but captures the important aspects of being fractionally more variable at lower flux and flattening at higher fluxes.


\subsection{Weakly stationary power spectra}
\label{sec:weak}

\begin{figure*}
\includegraphics[width=0.47\textwidth,angle=90]{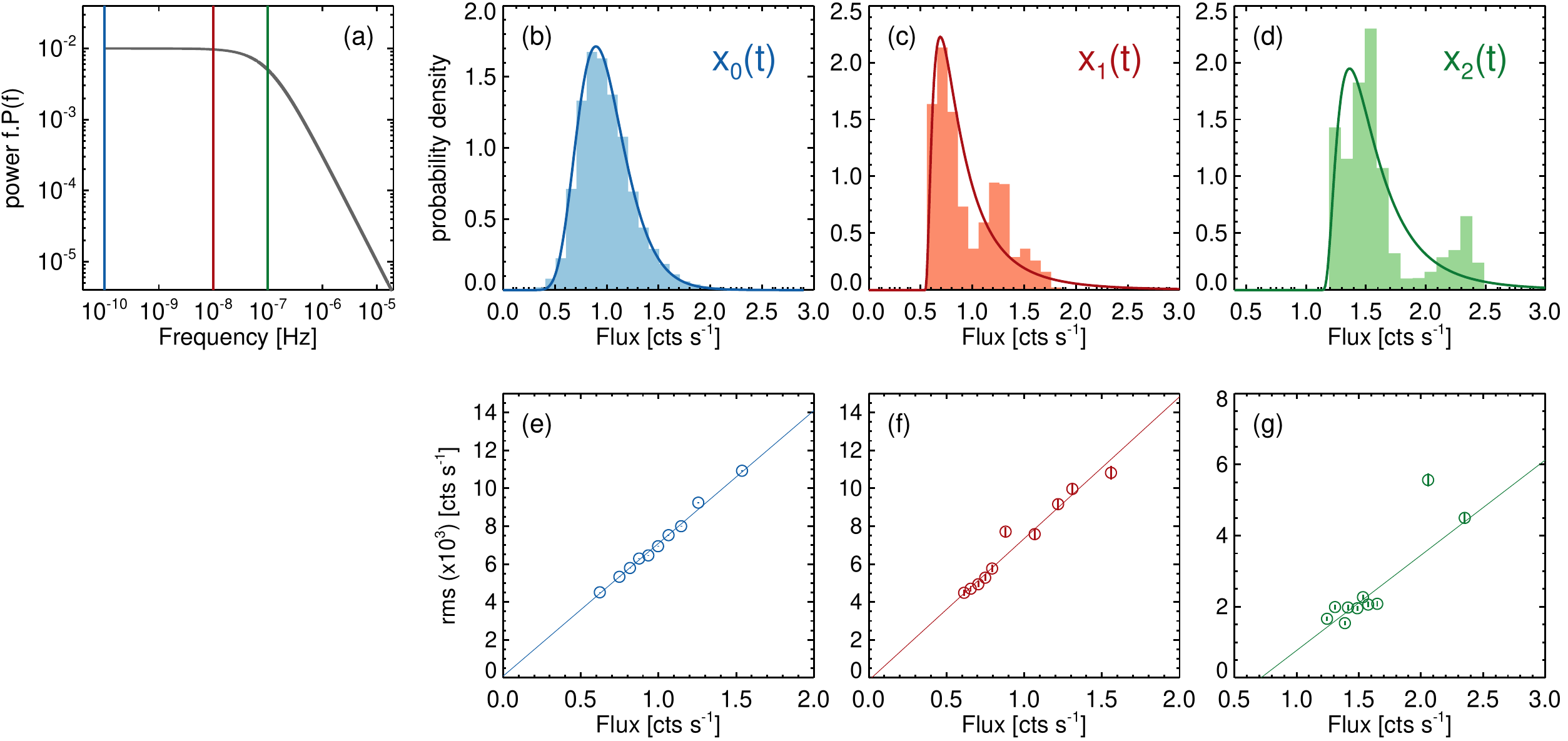}
\caption{The exponentiation model applied to \emph{weakly} stationary PSDs, which do not break to zero power at low frequencies (see text for details).  Panel (a) shows the rms normalised model PSD using a bending power-law function $\nu_{\rm b} = 1 \times 10^{-7}$\,Hz.  The full PSD provides the model light curve $x_{0}(t)$ with duration $T_{0} = 10^{10}$\,s (blue).  From this we take (non-overlapping) subsections to represent a `real' observations, $x_{1}(t)$ with $T_{1} = 10^{8}$\,s (red) and $x_{2}(t)$ with $T_{2} = 10^{7}$\,s (green).  Panels (b, c and d) show the flux distributions of the model light curves and best fitting log normal distribution.  Panels (e, f and g) show the rms-flux relation for each model light curve, calculated at frequencies above $\nu_{\rm b}$.  The solid lines are the best fitting linear relations to each model light curve.}
    \label{fig:weak}
\end{figure*}

In this section we look at the expected flux distributions and rms-flux relations from poorly sampled PSDs. In previous sections we have shown that in addition to the PSD, any valid variability model must be non-linear, and explain the flux distribution and rms-flux relation of the observations (see also \citetalias{uttleymchardyvaughan05} and \citetalias{uttleymchardyvaughan17}).  In some situations, observational constraints mean that the high frequency part of PSD is only observed, with the high frequency break often not detected.  This is particularly true for AGN as the characteristic variability timescales scale inversely with the mass of the accreting object, so are shifted to lower Fourier frequencies.   The light curves can then be described as being only \emph{weakly} stationary.  This may then lead to not sampling the full range of the flux distribution, and falsely detecting inconsistencies with the expectations of the exponentiation model.

We demonstrate this using a standard broadband PSD model formed from a bending power-law:

\begin{equation}
\label{eqn:bendpl}
   P(\nu) = \frac{N_{\rm b} \nu^{{\alpha}_{\rm low}}}{1 + (\nu / \nu_{\rm b})^{{\alpha}_{\rm low}-{{\alpha}_{\rm high}}}}
\end{equation}

\noindent where $\nu_{\rm b}$ is the bend frequency, ${\alpha}_{\rm low}$ and ${\alpha}_{\rm high}$ describe the slope below and above $\nu_{\rm bend}$ respectively, and $N_{\rm b}$ is the normalisation.  To represent typical observations we use $[\alpha_{\rm low},\nu_{\rm b},{\alpha}_{\rm high},N_{\rm b}] = [1.0, 10^{-7}, 2.5, 10^{-2}]$ and simulate model light curves in a similar manner to before.  We model the PSD down to $1 \times 10^{-10}$\,Hz, which is many orders of magnitude below the high frequency break.  The model light curve $x_{0}(t)$ has length $T = 1 \times 10^{10}$\,s and time resolution ${\rm dt}=10$\,s.  Following the exponentiation the light curves are rebinned to ${\rm dt}=500$\,s.

To represent `real' observations we take (non-overlapping) subsections from $x_{0}(t)$ with durations $T_{1} = 10^{8}$\,s: $x_{1}(t)$, and $T_{2} = 10^{7}$: $x_{2}(t)$.  These represent observations which measure the PSD down to just below and at $\nu_{\rm b}$, respectively. Fig.~\ref{fig:weak} (a) shows the model PSD and effective sampling for each of the model light curves.  Panels (b, c and d) show the flux distributions for each model light curve, along with the best fitting three-parameter lognormal distribution.  The full light curve $x_{0}(t)$ is well described by a lognormal.  The `real' light curves $x_{1}(t)$ and $x_{2}(t)$ clearly show a poor fit to this model, displaying a multimodal distribution which can not be captured by a simple lognormal.  Also apparent in these plots is how the means of the two sampled light curves are different to the original.

Fig.~\ref{fig:weak} panels (e, f and g) show the rms-flux relation for each model light curve, calculated at frequencies above $\nu_{\rm b}$.  The relation is calculated as before with $\ge 100$ estimates per flux bin.  As the rms-flux relation requires many estimates to average we use a different frequency range --- hence different segment length --- for each model light curve.  The rms is calculated in the range [$\nu_{\rm b}$,$\nu_{\rm Nyq}$]\,Hz, [$10^{-5}$,$\nu_{\rm Nyq}$]\,Hz and [$10^{-4}$,$\nu_{\rm Nyq}$]\,Hz, where $\nu_{\rm Nyq} = 1/2dt = 5 \times 10^{-3}$\,Hz is the Nyquist frequency, for $x_{0}(t)$, $x_{1}(t)$ and $x_{2}(t)$, respectively.  The rms-flux relation for $x_{0}(t)$ shows excellent agreement with a linear model.  There is some scatter, but this is expected for a stochastic process and using the simple exponentiation model (\citealt{vaughan03a}).  Despite being from a light curve with an intrinsic linear rms flux relation, $x_{1}(t)$ and $x_{2}(t)$ show a large amount of scatter and poor fit to a simple linear model.  

We note here that these subsections are random selections of the original realisation of the long light curve.  Different subsections show a variety of flux distributions with positive or negative skewness and bimodality.  A variety of rms-flux relations are also apparent, with the deviation from linear on average increasing as the light curve duration decreases.  The simulations presented here do not include additive Poisson noise (caused by measurement noise in real observations).  The affect of Poisson noise is to increase the width of the flux distribution and increase the scatter in the rms-flux relation.  However, this effect is marginal when the power on the observed timescales is much greater than the Poisson noise level, as is typically the case with real observations.

\section{Discussion}
\label{sec:disco}


Any model for the variability process in accreting systems needs to be able to produce observed flux distribution and rms-flux relation.  The simple exponentiation model is successful at describing the stationary light curves observed in accreting compact objects (\citetalias{uttleymchardyvaughan05}; \citetalias{uttleymchardyvaughan17}).  For the case of the AGN, \irasno, we have shown that the observed flux distribution and rms-flux relation can be reproduced using the exponential model.   The light curves are made up of piecewise stationary epochs which themselves have the typical properties.  The result of the sum of several non-stationary pieces is to distort the resulting observables.

This is important for understanding the source behaviour, as it means the underlying process is similar to that of other accreting sources, i.e. it has its origin in the accretion flow.  Some property of the standard process is changing with time to produce the observed light curve properties.  The question still remains of what is the origin of the non-stationarity in \irasno.  This is something which can be understood in MHD accretion disc simulations (e.g. \citealt{HoggReynolds16}), with the non-stationarity providing an extra handle for physical changes in the system.  To first order, the resulting flux distribution and rms-flux relation are not the result of some extrinsic modification of the standard variability process, for example variable absorption, as this is very different or non-existent in many other types of accreting compact objects. 
 
The changes in \iras correspond to changes in the PSD on $\sim 10$\,s in a $10 \Msun$ XRB, with the frequency range explored here equating to $\sim 1 - 100$\,Hz.  The shape of the broadband PSD means that \iras is a likely analogue of the transition state known as the very-high/intermediate state in BH XRBs (\citetalias{alston19a} and references therein).  Changes in the PSD and light curve behaviour on short timescales are seen in this transitional state (e.g. \citealt{belloni2000}; \citealt{nespoli03}).  In particular, quasi periodic oscillations are observed to appear and change frequency on timescales of $\sim 5$\,s (\citealt{nespoli03}).  These fast changes have been linked to radiation pressure driven instabilities in the accretion disc structure, expected when the accretion rate is a large fraction of the Eddington rate (e.g. \citealt{szuszkiewicz98}; \citealt{nayakshin2000}).  Understanding the observational properties of these non-stationary PSDs in XRBs may be possible with new instruments, such as NICER.

The model provides a nice description of the observed data in \irasno, but there are small differences.  These are expected given the exponentiation approach is very simplified, resulting in a trivial frequency coupling and a time-symmetric model light curve.  An improved approach is to include higher order measures such as the bicoherence (\citetalias{uttleymchardyvaughan05}; \citealt{maccaronecoppi02}).  In addition, a slight energy dependence to the PSD was found in \citetalias{alston19a}, which is only crudely replicated in the model PSDs and mean count rates here: the non-stationary behaviour may be different for neighbouring energy bands which is only approximated here using a broad energy band.  This is the focus of a further paper in a series on \irasno.

The non-linear rms-flux relation observed in \iras is an example of Simpson's paradox (\citealt{blyth72}; \citealt{wagner82}).  This is the phenomenon where a particular trend appears in separate groups, but disappears or even reverses when the groups are combined.  In the case of \irasno, the linear rms-flux trend is present over a short enough interval where the PSD is statistically stationary.  When estimates from a piecewise non-stationary light curve are combined, the individual linear trends are lost in the flux sorting, resulting in a distorted overall relation.

\citet{emmanoulopoulos13} presented an algorithm for producing simulated light curves with the same properties as the light curves for many types of astrophysical sources, including accreting compact objects.  In the special case where the desired output distribution of the simulation is lognormal, the simulated light curves also have a linear rms-flux relation, without the need for the exponentiation step.  This arises from the fact the \citet{timmerkonig95} method only produces Gaussian distributed light curves, so the lognormality needs to be imprinted.  However, like this method, the \citet{emmanoulopoulos13} algorithm also only generates stationary light curves for a given input PSD.  In order to produce the same results for \iras in Section~\ref{sec:iras} a piecewise non-stationary PSD approach will also have to be used.

\citet{smith18kepler} recently reported non-lognormal flux distributions in a sample of $\sim 20$ \textit{Kepler} optical AGN, with a variety of skewness and (bi)modality reported. This data also showed rms-flux relations inconsistent with a simple linear model.  The authors inferred that some different underlying variability process or accretion mode was at play in these sources.  However, the high-frequency PSD bend was poorly detected in these data, or at the lower end of the frequency window.  The results in Section~\ref{sec:weak} show how the measured flux distributions and rms-flux relations can differ from expectation, even when those light curves are sampled from a larger light curve which is lognormally distributed and have a linear rms-flux relation by design.  Caution must therefore be taken when making inferences about the underlying process when the PSD is only poorly sampled (i.e. is only weakly stationary).  This is something which can be trivially assessed using simulations similar to those presented here, but specific to the PSD and count rate of the data in question.  We note here that the optical emission from AGN can arise from multiple emission mechanisms and multiple emission regions, potentially providing a source of non-stationarity in the light curves.

\section{Conclusions}

We have demonstrated how the simple exponentiation model can successfully produce observed flux distribution and rms-flux relation.  The atypical results observed in the AGN, \irasno, are well described by the model when the PSD is \emph{piecewise} non-stationary.  Given the similarities in the variability process observed across accreting compact objects, it is likely that similar affects will be observed in further sources.  Understanding the cause of the non-stationarity will then provide further insights into the accretion process.  There will indeed be some deviations from the model due to e.g. changes in source state, long term accretion rates, absorption and obscuration events.  These can of course all be accounted for in the simulation process.

With longer baseline AGN variability studies becoming available it will be tempting to use these in a similar manner to what has classically been achieved with X-ray timing studies (see e.g. \citealt{vaughan16}).  Before any claims of disagreements with simple model can be made, it is important to check your results with simulated light curves with the correct intrinsic properties to see what the flux distribution and rms-flux relation should look like for the observed PSD.

\section*{Acknowledgments}

WNA acknowledges discussions on accretion disc simulations with Chris Reynolds.  WNA acknowledges support from the European Research Council through Advanced Grant 340492, \textit{`Feedback'}.  This paper is based on observations obtained with \xmm, an ESA science mission with instruments and contributions directly funded by ESA Member States and the USA (NASA). 




\bibliographystyle{mnras}
\bibliography{nonstat} 








\bsp	
\label{lastpage}
\end{document}